\documentclass[aps,prl,twocolumn,groupedaddress]{revtex4}
\usepackage{amsmath}
\usepackage{amsfonts, amsthm, amssymb}

\begin{document}

\title{Spin-Josephson effect in antiferromagnetic tunnel junctions}
\author{D. Chass\'e and A.-M.S. Tremblay }
\affiliation{D\'epartment de physique and RQMP, Universit\'e de Sherbrooke, Sherbrooke,
Qu\'ebec, Canada J1K 2R1.}
\date{\today }

\begin{abstract}
We study Josephson-like phenomena in a tunnel junction between two itinerant antiferromagnets.
We find at the mean-field level an equilibrium current of the staggered magnetic moment through
the junction that is proportional to the normal state conductance and to $\mathbf{S}_{L}$$\times$$\mathbf{S}_{R}$
where $\mathbf{S}_{L}$ and $\mathbf{S}_{R}$ are the staggered magnetic moments on either sides. Microscopically,
this effect comes from the coherent tunneling of spin-one charge-zero particle-hole pairs with a net wave vector equal to
the antiferromagnetic one. We explain similarities and differences with the standard DC and AC Josephson effects.
\end{abstract}


\maketitle










One of the most striking manifestations of superconductivity is the
Josephson effect, which consists in the coherent tunneling of Cooper pairs across a junction when both sides are superconductors. Interestingly, this occurs even though the Hamiltonian contains only single-electron tunneling.
 The resulting current is proportional to $\sin (\varphi )$,
where $\varphi $ is the difference between the phases of the two
superconducting order parameters. In the presence of a constant potential
difference, gauge invariance requires the phase difference to increase
linearly with time, leading to an alternating current whose frequency
depends only on universal physical constants and not on any material
parameter.

Superconductivity is just one example of coherence and spontaneous symmetry
breaking. The order parameter is a complex number so the broken symmetry is $%
U(1)$. The superconducting state selects a phase and the Josephson effect
arises from the tendency to make the phase uniform across the tunnel
junction.

Since there are many other types of order and corresponding broken
symmetries, the question of the analog of the Josephson effect in such cases
arises naturally \cite{Paranjape}. Indeed, one should expect differences
in order parameters across a junction to lead to a coherent tunneling of the condensed objects
that exist in the broken symmetry state. This possibility is especially relevant in the context where junctions
between magnetic materials are of utmost importance for spintronics. In fact,
theoretical predictions have been made recently concerning the possible
existence of a Josephson-like equilibrium spin current in ferromagnetic (FM)
tunnel junctions in analogy with superconducting junctions ~\cite{Lee,
Nogueira, wang}. FM long range order is a realization of
spontaneous $SO(3)$ symmetry breaking and an equilibrium spin current would
result from the exchange coupling between the magnetic moments in the two
leads, which favors alignment of order parameters. From a Ginzburg-Landau
point of view, there is, in the functional, a term proportional to $\mathbf{M%
}_{L}\cdot \mathbf{M}_{R}$ where $\mathbf{M_{L}}$ and $\mathbf{M_{R}}$ are,
respectively, the magnetic moments on the left and on the right of the
junction. The Heisenberg equations of motion thus lead to $\tfrac{d\mathbf{%
M_{L}}}{dt}\sim \mathbf{M_{L}}\times \mathbf{M_{R}}$, corresponding to a
spin current.


Josephson-like phenomena between antiferromagnets (AF) are interesting for
several reasons. In the context of spintronics, it has been shown experimentally that the absence of a net angular
momentum in AF results in orders of magnitude faster spin
dynamics than in FM, which could expand the now limited set of applications
for AF materials \cite{Kimel}.
More generally, in AF the N\'{e}el order parameter breaks
both lattice translation symmetry and $SO(3)$ spin rotation symmetry so the
situation is less straightforward than in FM from a mathematical point of view. In addition,
AF are often close to superconducting phases, as is found in
heavy fermions, high-temperature superconductors and layered organic
superconductors. Understanding Josephson-like phenomena in AF
is the first step towards more general studies with coexisting
antiferromagnetic and superconducting order parameters. Generalized
Josephson effects may help identify homogeneous coexistence in real materials. Studies
along these lines have recently appeared for ferromagnetism coexisting with
other types of order \cite{Sudbo:2007}.

In this paper, we consider a microscopic model for single electron tunneling
between itinerant AF. This microscopic calculation leads to an
explicit expression for the analog of the critical current and its
temperature dependence. In addition, we show that (a) Cooper-pair tunneling
is replaced by tunneling of a spin-one neutral particle-hole pair (b) time
dependence introduced by external magnetic fields resemble the AC Josephson
effect but there are many differences because of the non-abelian nature of
the problem and because spins do not couple to the gauge field but direcly
to the magnetic field. Finally, we briefly discuss how such effects can in principle
be observed experimentally.

\textit{The model} The Hamiltonian for a tunneling junction consisting of
two leads of an AF material and an insulating barrier
between them reads
\begin{equation}
H=H_{L}(c^{\dag}_{\mathbf{k}\sigma},c_{\mathbf{k}\sigma})+H_{R}(d^{\dag}_{%
\mathbf{q}\sigma},d_{\mathbf{q}\sigma})+H_{T} \;,
\end{equation}
\noindent where $H_{L(R)}$ is the Hamiltonian of the left (right) AF, $H_{T}$
is the tunneling part connecting the two leads, and $c^{\dag}_{\mathbf{k}%
\sigma}(c_{\mathbf{k}\sigma})$ and $d^{\dag}_{\mathbf{q}\sigma}(d_{\mathbf{q}%
\sigma})$ are the fermion creation (annihilation) operators of the left and
right leads, respectively. In the following discussion the quantum numbers $%
\mathbf{k}$ and $\mathbf{q}$ will also denote implicitly the left and right
lead.

We model the AF on each side of the junction by a one-band Hubbard
Hamiltonian treated in the Hartree-Fock approximation for a static
spin-density wave (SDW) with wave-vector $\mathbf{Q}$. Without loss of
generality, we assume the SDW mean field to be polarized along the spin
quantization axis. Following Ref.~\cite{schrieffer}, we write
\begin{equation}  \label{HF_Hubbard}
\hat{H}_{L}=\sum_{\mathbf{k}\alpha}\epsilon_{k}c^{\dag}_{\mathbf{k}\alpha}c_{\mathbf{k}\alpha}
-\dfrac{US}{2}\sum_{\mathbf{k}\alpha\beta}c^{\dag}_{\mathbf{k}+\mathbf{Q}%
\alpha}\sigma^{3}_{\alpha\beta} c_{\mathbf{k}\beta} \;,
\end{equation}
\noindent where $\epsilon _{k}$ is the band dispersion, $U$ the interaction
strength, $\sigma ^{3}$ the third Pauli matrix, while the order parameter $S$ is defined by $(1/N)\big\langle\sum_{\mathbf{k}\alpha \beta
} c_{\mathbf{k}+\mathbf{Q}\alpha }^{\dag }\sigma^{3}_{\alpha \beta }c_{%
\mathbf{k}\beta }
\big\rangle $ with $N$ the number of sites. This one-body Hamiltonian can be diagonalized by the
Bogoliubov transformation
\begin{equation}
\begin{aligned}
\gamma _{\mathbf{k}\alpha}^{c} &=u_{\mathbf{k}}c_{\mathbf{k}\alpha}+v_{\mathbf{k}%
}\sum_{\beta }(\sigma ^{3})_{\alpha\beta }c_{\mathbf{k}+\mathbf{Q}\beta }\;,
\label{Bogoliubov} \\
\gamma _{\mathbf{k}\alpha}^{v} &=v_{\mathbf{k}}c_{\mathbf{k}\alpha}-u_{\mathbf{k}%
}\sum_{\beta }(\sigma ^{3})_{\alpha\beta }c_{\mathbf{k}+\mathbf{Q}\beta }\;.
\end{aligned}
\end{equation}
\noindent To avoid double counting, $\mathbf{k}$ is restricted to the
magnetic zone. The superscripts $c$ and $v$ refer to the conduction and the
valence bands split by the exchange Bragg scattering from the SDW. For
simplicity, we assume perfect nesting $\epsilon _{\mathbf{k}}=-\epsilon _{%
\mathbf{k}+\mathbf{Q}}$. In this case, the coefficients of the
transformation are given by $u_{\mathbf{k}}^{2}=\big[\tfrac{1}{2}\big(%
1+\epsilon _{\mathbf{k}}/E_{\mathbf{k}}\big)\big]$, $v_{\mathbf{k}}^{2}=\big[%
\tfrac{1}{2}\big(1-\epsilon _{\mathbf{k}}/E_{\mathbf{k}}\big)\big]$, $E_{%
\mathbf{k}}^{2}=(\epsilon _{\mathbf{k}}^{2}+\Delta ^{2})$, where $\Delta
=-US/2$ is the SDW gap parameter \cite{half}. The diagonalized Hamiltonian
is given by $H=\sum_{\mathbf{k}\alpha}^{\ast }E_{\mathbf{k}}\big(\gamma _{%
\mathbf{k}\alpha}^{\dag c}\gamma _{\mathbf{k}\alpha}^{c}-\gamma _{\mathbf{k}%
\alpha}^{\dag v}\gamma _{\mathbf{k}\alpha}^{v}\big)$ where $\sum_{\mathbf{k}}^{\ast
} $ means that the sum extends over the magnetic zone. The single-particle
energy spectrum is given by $\pm E_{\mathbf{k}}$ and the SDW ground state
for a half-filled band is defined by $\gamma _{\mathbf{k}\alpha}^{\dag v}|\Omega
\rangle =\gamma _{\mathbf{k}\alpha}^{c}|\Omega \rangle =0$, which corresponds
to:
\begin{align}
\left\vert \Omega \right\rangle & =\Pi _{\mathbf{k}\alpha }^{\ast }(v_{%
\mathbf{k}}c_{\mathbf{k}\alpha }^{\dagger }-u_{\mathbf{k}}\sum_{\beta }c_{%
\mathbf{k+Q}\beta }^{\dagger }\sigma _{\beta \alpha }^{3})\left\vert
0\right\rangle  \label{GS_standard} \\
& =\Pi _{\mathbf{k}\alpha }^{\ast }(v_{\mathbf{k}}-u_{\mathbf{k}%
}\sum_{\beta }c_{\mathbf{k+Q}\beta }^{\dagger }\sigma _{\beta \alpha }^{3}c_{%
\mathbf{k}\alpha })c_{\mathbf{k}\alpha }^{\dagger }\left\vert 0\right\rangle .
\label{GS_BCS}
\end{align}%
The last form makes the analogies with the BCS ground state clear. For
example, there exists Andreev-like reflections at AF-N
interfaces~\cite{Bobkova:2005}. The ground state contains coherent
particle-hole pairs. To clarify this, we perform a particle-hole
transformation for states that are in the first magnetic-Brillouin zone.
Recall that destroying an electron in a state, creates a hole in the
corresponding time-reversed state. For a spinor this can be achieved by $c_{%
\mathbf{k}\uparrow }\rightarrow h_{-\mathbf{k}\downarrow }^{\dagger }$ and $%
c_{\mathbf{k}\downarrow }\rightarrow -h_{-\mathbf{k}\uparrow }^{\dagger }.$
The ground state then takes the form
\begin{equation}
\left\vert \Omega \right\rangle =\Pi _{\mathbf{k}}^{\ast }(v_{\mathbf{k}}-u_{%
\mathbf{k}}c_{\mathbf{k+Q}\uparrow }^{\dagger }h_{-\mathbf{k}\downarrow
}^{\dagger })(v_{\mathbf{k}}-u_{\mathbf{k}}c_{\mathbf{k+Q}\downarrow
}^{\dagger }h_{-\mathbf{k}\uparrow }^{\dagger })\left\vert 0\right\rangle
_{h}
\end{equation}%
\noindent where $c_{\mathbf{k+Q}\alpha }\left\vert 0\right\rangle _{h}=0$ and $h_{\mathbf{k}%
\alpha }\left\vert 0\right\rangle _{h}=0$. The particle-hole pair
is in a triplet state with vanishing net spin projection along the
quantization axis, has no charge and has a wave vector equal to the
antiferromagnetic wave vector. In the case of a FM, that wave
vector would vanish.

Since we choose a different quantization axis in the two AF, we need to
include a unitary transformation in spin space denoted by $\mathbf{U}(\theta
,\phi )$ to account for the fact that a $\uparrow $ spin on one side of the
junction is not the same as a $\uparrow $ spin on the other side of the
junction. 
Since the spin quantization axes are taken along the direction of the staggered magnetic moment $\mathbf{S}$
on each side, the angles $(\theta ,\phi )$ correspond to the orientation of $\mathbf{S}_{R}$
of the right AF expressed in the coordinate system of the left
side of the junction; in Cartesian coordinate $\mathbf{S}_{L}=|\mathbf{S}_{L}|(0,0,1)$ and $\mathbf{S}_{R}=|\mathbf{S}_{R}|(\sin
\theta \cos \phi ,\sin \theta \sin \phi ,\cos \theta )$. The tunneling
Hamiltonian then reads $\hat{H}_{T}=(1/N)\sum_{\mathbf{k}\mathbf{q}\sigma \sigma
^{\prime }}\big(\mathbf{U}_{\sigma \sigma ^{\prime }}t_{\mathbf{k}\mathbf{q}%
}c_{\mathbf{k}\sigma }^{\dag }d_{\mathbf{q}\sigma ^{\prime }}+h.c.\big)$.
The spin flip terms come purely from the choice of different quantization
axes on the left and on the right. There is no real spin flip in the
tunneling process. Note also that the generators of the unitary
transformation $\mathbf{U}(\theta ,\phi )$ are non-abelian gauge fields that
are the analog of the electromagnetic gauge potentials in the ordinary
Josephson effect. The former do not couple to the electromagnetic fields and
they do not have a dynamics independent of that of the spins.

\textit{Derivation of the equations of motion}. The staggered magnetic moment operator
in the left lead is $\mathbf{\hat{S}}_{L}=(\hbar /2)\sum_{\mathbf{k}\alpha \beta
}c_{\mathbf{k}+\mathbf{Q}\alpha }^{\dag }\mathbf{\mathbf{\sigma }}_{\alpha \beta }c_{%
\mathbf{k}\beta }$ where $\mathbf{\sigma }$ is a vector of Pauli matrices.
In the broken symmetry state, the time evolution of $\mathbf{\hat{S}}_{L}$ due to $%
H_{L}$ is negligible. Since we also have $[\mathbf{\hat{S}}_{L},\hat{H}_{R}]=0$ we find $%
d\mathbf{\hat{S}}_{L}/dt=(1/i\hbar )[\mathbf{\hat{S}}_{L},\hat{H}_{T}]$, and thus $d
\mathbf{\hat{S}}_{L}/dt=-(i/2N)\sum_{\mathbf{k}\mathbf{q}}\sum_{\alpha \beta
\delta }\big(\mathbf{\sigma }_{\alpha \beta }\mathbf{U}_{\beta \delta }\,t_{%
\mathbf{k}\mathbf{q}}c_{\mathbf{k}+\mathbf{Q}\alpha }^{\dag }d_{\mathbf{q}%
\delta }-h.c.\big)$ whose average $\dot{\mathbf{S}}_{L}(t) \equiv \langle d\mathbf{\hat{S}}_{L}/dt \rangle$ is given by
\begin{equation}
\dot{\mathbf{S}}_{L}(t) =\dfrac{1}{N}\sum_{\mathbf{k}\mathbf{q}%
}\sum_{\alpha \beta \delta }\text{Im}\Big[\mathbf{\sigma }_{\alpha \beta }%
\mathbf{U}_{\beta \delta }\,t_{\mathbf{k}\mathbf{q}}\big\langle c_{\mathbf{k}+%
\mathbf{Q}\alpha }^{\dag }(t)d_{\mathbf{q}\delta }(t)\big\rangle \Big]\;,
\label{Kubo1}
\end{equation}%
\noindent where $\langle ...\rangle $ is the thermal statistical average with the full
density matrix. Performing first order perturbation theory using $\hat{H}_{T}$ as
the perturbation, one obtains
\begin{equation}
\big\langle c_{\mathbf{k}+\mathbf{Q}\alpha }^{\dag }d_{\mathbf{q}\delta
}\big\rangle =-\dfrac{i}{\hbar}\int_{-\infty }^{t}dt^{\prime }\big\langle \big[c_{\mathbf{k}+%
\mathbf{Q}\alpha }^{\dag }(t)d_{\mathbf{q}\delta }(t),\hat{H}_{T}(t^{\prime })\big]%
\big\rangle _{0}\;,  \label{Kubo2}
\end{equation}%
\noindent where the average $\langle ...\rangle _{0}$ is computed with $H_{0}=\hat{H}_{L}+\hat{H}_{R}$
(the unperturbed part of $\hat{H}$). The operators on the right are in the
interaction representation.

Once the commutator is evaluated in Eq.~(\ref{Kubo2}),
one of the factorizations of the
four-point correlation function involves products of correlation functions
on the left and on the right leads such as $\langle c_{\mathbf{k}+\mathbf{Q}\alpha
}^{\dag }(t)c_{\mathbf{k}\alpha }(t^{\prime })\rangle _{0}\langle d_{\mathbf{%
q}+\mathbf{Q}\delta }(t)d_{\mathbf{q}\delta }^{\dag }(t^{\prime })\rangle
_{0}$. Such correlation functions would vanish in a normal paramagnetic
state. They are non-zero because of the broken symmetry. They represent
interference in the tunneling process between momentum $\mathbf{k+Q}$ spin
up particles and momentum $-\mathbf{k}$ spin-down holes, in other words
tunneling of charge zero spin one $S^{z}=0$ coherent particle-hole pairs that have finite
momentum and are present in the ground-state Eq.(\ref{GS_BCS}). In the case
of the ordinary Josephson effect, one would find terms such as $\langle c_{%
\mathbf{k}\sigma }^{\dag }(t)c_{-\mathbf{k}-\sigma }^{\dag }(t^{\prime
})\rangle _{0}\langle d_{\mathbf{q}\sigma }(t)d_{-\mathbf{q}-\sigma
}(t^{\prime })\rangle _{0}$ that represent tunneling of coherent Cooper
pairs.

In order to compute the averages $\langle...\rangle_0$ in the broken symmetry states, we
invert the Bogoliubov transformation Eq.~(\ref{Bogoliubov}). Assuming $t_{\mathbf{k}
\mathbf{q}}=t_{\mathbf{k}\mathbf{q+Q}}=t_{\mathbf{k+Q}\mathbf{q}}=t_{\mathbf{%
k+Q}\mathbf{q+Q}}$, we find

\begin{equation}
\sum_{\mathbf{k}\mathbf{q}}t_{\mathbf{k}\mathbf{q}}c^{\dag}_{\mathbf{k}+\mathbf{Q}\alpha} d_{\mathbf{%
q}\delta}=\sum^{*}_{\mathbf{k}\mathbf{q}}\sum_{i,j \in
\{c,v\}}t_{\mathbf{k}\mathbf{q}}(\Gamma^{\alpha\delta}_{\mathbf{k}\mathbf{q}})_{ij} \gamma^{i\dag}_{%
\mathbf{k}\alpha}\gamma^{j}_{\mathbf{q}\delta},
\end{equation}
where we defined
\begin{equation}
\begin{aligned} &(\Gamma^{\alpha\delta}_{\mathbf{k}\mathbf{q}})_{cc}\equiv
(u_{\mathbf{k}}u_{\mathbf{q}}+
\sigma^{3}_{\alpha\alpha}v_{\mathbf{k}}u_{\mathbf{q}}+\sigma^{3}_{\delta%
\delta}u_{\mathbf{k}}v_{\mathbf{q}} +
\sigma^{3}_{\alpha\alpha}\sigma^{3}_{\delta\delta}
v_{\mathbf{k}}v_{\mathbf{q}} )\\
&(\Gamma^{\alpha\delta}_{\mathbf{\mathbf{k}}\mathbf{q}})_{cv}\equiv
(v_{\mathbf{k}}u_{\mathbf{q}}-
\sigma^{3}_{\alpha\alpha}u_{\mathbf{k}}u_{\mathbf{q}}+\sigma^{3}_{\delta%
\delta}v_{\mathbf{k}}v_{\mathbf{q}} -
\sigma^{3}_{\alpha\alpha}\sigma^{3}_{\delta\delta}
u_{\mathbf{k}}v_{\mathbf{q}} )\\
&(\Gamma^{\alpha\delta}_{\mathbf{k}\mathbf{q}})_{vc}\equiv
(u_{\mathbf{k}}v_{\mathbf{q}}+
\sigma^{3}_{\alpha\alpha}v_{\mathbf{k}}v_{\mathbf{q}}-\sigma^{3}_{\delta%
\delta}u_{\mathbf{k}}u_{\mathbf{q}} -
\sigma^{3}_{\alpha\alpha}\sigma^{3}_{\delta\delta}
v_{\mathbf{k}}u_{\mathbf{q}} )\\
&(\Gamma^{\alpha\delta}_{\mathbf{k}\mathbf{q}})_{vv}\equiv
(v_{\mathbf{k}}v_{\mathbf{q}}-
\sigma^{3}_{\alpha\alpha}u_{\mathbf{k}}v_{\mathbf{q}}-\sigma^{3}_{\delta%
\delta}v_{\mathbf{k}}u_{\mathbf{q}} +
\sigma^{3}_{\alpha\alpha}\sigma^{3}_{\delta\delta}
u_{\mathbf{k}}u_{\mathbf{q}} )\;. \end{aligned}
\end{equation}

Similarly, let $\tilde{H}_{T}$ denote the part of $\hat{H}_{T}$ that does not
commute with $c^{\dag}_{\mathbf{k}+\mathbf{Q}\alpha} d_{\mathbf{q}\delta}$.
It can be written as $\tilde{H}_{T}=(1/N)\sum^{*}_{\mathbf{k}\mathbf{q}%
\sigma\delta^{\prime}}\sum_{ij}\mathbf{U}^{*}_{\sigma\delta^{\prime}}t^{*}_{%
\mathbf{k}\mathbf{q}} (\Gamma^{\sigma\delta^{\prime}}_{\mathbf{k}\mathbf{q}%
})_{ij} \gamma^{j\dag}_{\mathbf{q}\delta^{\prime}}\gamma^{i}_{\mathbf{k}%
\sigma}$. 
Substituting these expressions into Eq. (\ref{Kubo2}), one finds
\begin{widetext}
\begin{equation}\label{Js_G_omega}
\begin{aligned}
                \dot{\mathbf{S}}_{L}(t)
	&=\dfrac{1}{N^2}\sum^{*}_{\mathbf{k}\mathbf{q}}\sum_{\alpha\beta\delta}\sum_{\sigma\delta'}\text{Im}\Bigg[-\dfrac{i}{\hbar}
	|t_{\mathbf{k}\mathbf{q}}|^{2}\vec{\sigma}_{\alpha\beta}
	\mathbf{U}_{\beta\delta}\mathbf{U}^{*}_{\sigma\delta'}\int dt' e^{-0^{+}(t-t')}\times\\
	&\qquad\qquad\qquad\qquad\qquad\qquad\sum_{ij}
	(\Gamma^{\alpha\delta}_{\mathbf{k}\mathbf{q}})_{ij}(\Gamma^{\sigma\delta'}_{\mathbf{k}\mathbf{q}})_{ij}
	\Big(\mathcal{G}^{i<}_{\mathbf{k}\sigma\alpha}(t'-t)
	\mathcal{G}^{j>}_{\mathbf{q}\delta\delta'}(t-t') -
	\mathcal{G}^{i>}_{\mathbf{k}\sigma\alpha}(t'-t)\mathcal{G}^{j<}_{\mathbf{q}\delta\delta'}(t-t')\Big)\Bigg]
	\end{aligned}
	\end{equation}
\end{widetext}
where $\mathcal{G}^{i<(>)}_{\mathbf{k}(\mathbf{q})}$ are the Keldysh Green
functions in the left (right) lead. Their definitions are $\mathcal{G}^{i<}_{%
\mathbf{k}(\mathbf{q}),\alpha\beta} (t,t^{\prime})=i\langle\gamma^{i\dag}_{%
\mathbf{k}(\mathbf{q})\beta}(t^{\prime})\gamma^{i}_{\mathbf{k}(\mathbf{q}%
)\alpha}(t) \rangle$ and $\mathcal{G}^{i>}_{\mathbf{k}(\mathbf{q}%
),\alpha\beta} (t,t^{\prime})=-i\langle\gamma^{i}_{\mathbf{k}(\mathbf{q}%
)\alpha}(t)\gamma^{i\dag}_{\mathbf{k}(\mathbf{q})\beta}(t^{\prime}) \rangle$%
, respectively. Explicitly,
\begin{align*}
G^{i >}_{\mathbf{k}\sigma\sigma^{\prime}}(t^{\prime}-t)&= -i(1-f(E^{i}_{k}))%
\exp[-i E^{i}_{k} (t^{\prime}-t)/\hbar]\delta_{\sigma\sigma^{\prime}}\;, \\
G^{i <}_{\mathbf{k}\sigma\sigma^{\prime}}(t^{\prime}-t)&= i f(E^{i}_{k}) \exp%
[-i E^{i}_{k} (t^{\prime}-t)/\hbar]\delta_{\sigma\sigma^{\prime}} \;,
\end{align*}
where $f$ is the Fermi function. Eq.~(\ref{Js_G_omega}) for the staggered magnetic moment current through a
tunnel junction is general. A bias could be included. We assume that there is no bias so there is no incoherent
single-particle tunneling across the antiferromagnetic gap. Integrating over
$t-t^{\prime }$ and performing the spin sum in (\ref{Js_G_omega}), one finds
\begin{equation}
\dot{\mathbf{S}}_{L}=I_{c}\,\hat{\mathbf{s}}_{R}\times
\hat{\mathbf{s}}_{L}\;,  \label{mainresult1}
\end{equation}%
\noindent where $\hat{\mathbf{s}}_{L(R)}=\mathbf{S}_{L(R)}/|\mathbf{S}_{L(R)}|$ and
\begin{align}
\label{mainresult2}
I_{c}& =\dfrac{8\Delta _{L}\Delta _{R}}{N^2}P\sum_{kq}^{\ast }|t_{\mathbf{kq}}|^{2}\dfrac{f(E_{\mathbf{k}})-f(-E_{\mathbf{k}})}{E_{\mathbf{k}}(E_{%
\mathbf{k}}^{2}-E_{\mathbf{q}}^{2})}  
\end{align}%
with $P$ indicating principal part. A similar expression is found for the equilibrium spin
current in the case of ferromagnetic tunnel junctions. Note that the sine
function present in the standard Josephson case is replaced here by a cross
product, which is a direct consequence of the vectorial nature of the order
parameter.

For a symmetrical junction ($\Delta_{L}=\Delta_{R}$), the same assumptions
and procedure as Ref.~\cite{ambegaokar} lead to the following analytical
result
\begin{equation}
I_{c}=\dfrac{h}{e^{2}}R^{-1}\Delta (T)\tanh (\tfrac{1}{2}\beta \Delta (T))\; ,
\end{equation}
\noindent where $R=\hbar/\left( 4\pi e^{2}D^2\left\vert t\right\vert ^{2}\right) $ is
the (zero-temperature) normal-state resistance of the junction with $D$ the density of state,
which is assumed to be a constant. This expression for the
temperature dependance of the critical current has the same form as that
obtained by Ambegaokar and Baratoff \cite{ambegaokar} for a BCS
superconductor, which is not surprising given the formal analogies \cite{Over}.

By symmetry, the time derivative of the stagerred magnetic moment on the right
lead can be obtained by interchanging the $L$ and $R$ indices in Eq.~(\ref%
{mainresult1}). As a consequence, the staggerred magnetic moments of the two
AF precess about their (constant) sum $\mathbf{S}_{L}+\mathbf{S}_{R}$ at a frequency $\omega _{0}=I_{c}|\mathbf{S}_{L}+\mathbf{S}_{R}|/|\mathbf{S}_{L}||\mathbf{S}_{R}|$. In the limit where 
$\mathbf{S}_{L}$ and $\mathbf{S}_{R}$ are nearly colinear, the effect should therefore lead to a uniform shift of order $\omega _{0}$ in the antiferromagnetic resonance frequencies of each AF \cite{Kittel}.
With $|\tfrac{2\Delta }{S}|=U=2\text{eV} $ and a normal state
conductance of the order of the conductance quantum, $R^{-1}=2e^{2}/h$, one
finds a value of $\omega _{0}$ of order $10^{14}$ Hz, in other words in the visible.
This frequency would be higher than the single particle gap and so would
lead to much damping. Resistances that are many orders of magnitude larger
are thus needed to bring its value down.

\textit{AC spin-Josephson effect}. In the ordinary Josephson effect, the electromagnetic gauge
potentials enter directly in the argument of the sine function. The present
case is different. Each magnetic moment associated with a spin couples to
the magnetic field through the Zeeman term ($H_{Z}=-g\mu _{B}\mathbf{B\cdot S}$)
where $g$ is the gyromagnetic ratio and $\mu _{B}$ the Bohr magneton (in insulators we can
neglect terms coming from orbital motion).

Considering magnetic fields $\mathbf{B}_{L}$ and $\mathbf{B}_{R}$ applied
respectively to the left- and right-hand sides of the junction, the
Heisenberg equations of motion lead to the following equations of motion for
the order parameters
:
\begin{align}
\dot{\mathbf{S}}_{L} & =-g\mu _{B}\mathbf{B}_{L}\times\mathbf{S}_{L} +I_{c}\,\hat{\mathbf{s}}_{R}\times
\hat{\mathbf{s}}_{L}, \nonumber \\
\dot{\mathbf{S}}_{R} & =-g\mu _{B}\mathbf{B}_{R}\times\mathbf{S}_{R} +I_{c}\,\hat{\mathbf{s}}_{L}\times
\hat{\mathbf{s}}_{R}\;.
\label{dS/dt_AC}
\end{align}
The first term on the right side of the equality is merely the contribution of $H_{Z}$ to the Heisenberg equation of motion. The second term is the tunneling contribution and has exactly the same form as the one we have already computed in the zero-field case, namely Eq.(\ref{mainresult1}) and (\ref{mainresult2}). To see this, return to Eq.(\ref{Kubo2}). The only terms from $H_{T}(t')$ that give a non-zero contribution to the average are of the form $d^{\dag}_{\mathbf{q}\delta'}(t')U^{\dag}_{\delta'\sigma'}(t')c_{\mathbf{k}\sigma'}(t')$. Since these operators are in the interaction representation and $[H_{0},H_{Z}]=0$, it is possible to write $d^{\dag}_{\mathbf{q}\delta'}(t')=d^{\dag}_{\mathbf{q}\delta}(\tau')\Lambda^{R\dag}_{\delta\delta'}$ and $c_{\mathbf{k}\sigma'}(t')=\Lambda^{L}_{\sigma'\sigma}c_{\mathbf{k}\sigma}(\tau')$ where $\Lambda^{R(L)}_{\sigma'\sigma}=\text{exp}[ig\mu_{B}\mathbf{B}_{R(L)}\cdot\mathbf{\sigma}(t'-t)/\hbar]_{\sigma'\sigma}$ and $\tau'$ stands for the time evolution due to $H_{0}$. Similarly, one can show that the unitary transformation $U^{\dag}_{\delta'\sigma'}(t')$ can be written as $\Lambda^{R}_{\delta'\delta}U^{\dag}_{\delta\sigma}(t)\Lambda^{L\dag}_{\sigma\sigma'}$. We then obtain $d^{\dag}_{\mathbf{q}\delta'}(t')U^{\dag}_{\delta'\sigma'}(t')c_{\mathbf{k}\sigma'}(t')= d^{\dag}_{\mathbf{q}\delta}(\tau')U^{\dag}_{\delta\sigma}(t)c_{\mathbf{k}\sigma}(\tau')$, i.e all spin indices are at time $t$. Therefore, it is possible to factor the unitary transformation $U^{\dag}_{\delta\sigma}(t)$ out of the integral over $t'$, so that the rest of the calculation just follows the same path as in the zero-field case, thus leading to Eq. (\ref{dS/dt_AC}).

If $\mathbf{B}_{L}=\mathbf{B}_{R}=\mathbf{B}$, Eq.(\ref{dS/dt_AC}) implies that in the rotating frame $\mathbf{\hat{u}}=\{\mathbf{\hat{x}}
^{\prime },\mathbf{\hat{y}}^{\prime },\mathbf{\hat{z}}^{\prime }\}$ defined by
$d\mathbf{\hat{u}}/dt=-g\mu _{B}
\mathbf{B}\times \mathbf{\hat{u}}$, $\mathbf{S}_{L}$ and $\mathbf{S}_{R}$ still precess about their (constant) sum $\mathbf{\Sigma}\equiv\mathbf{S}_{L}+\mathbf{S}_{R}$ at a frequency $\omega_{0}$. Returning to the static frame, Eq.(\ref{dS/dt_AC}) gives $d \mathbf{\Sigma}/dt=-g\mu _{B}
\mathbf{B}\times\mathbf{\Sigma}$ so that $\mathbf{S}_{L}$ and $\mathbf{S}_{R}$ undergo a motion of double precession.

In the ordinary Josephson effect, a constant electric potential difference $V$
leads to a phase difference $\varphi $ that depends linearly on time ($\dot{\varphi}=2eV/\hbar $). In the present case, an analog is found by
computing the time dependence of the relative orientation $\vartheta $
between $\mathbf{S}_{L}$ and $\mathbf{S}_{R}$. Using Eq. (\ref{dS/dt_AC}) to compute $d(\cos
\vartheta )/dt=d(\hat{\mathbf{s}}_{L}\cdot \hat{\mathbf{s}}_{R})/dt$, we obtain
\begin{equation}
\dot{\vartheta}(t)=-g\mu _{B}\delta \mathbf{B}\cdot\hat{\mathbf{e}}(t) ,
\end{equation}%
where $\delta\mathbf{B}\equiv\mathbf{B}_{L}-\mathbf{B}_{R}$ and $\hat{\mathbf{e}}(t)\equiv\hat{\mathbf{s}}_{L}(t)\times
\hat{\mathbf{s}}_{R}(t)$. As in the ordinary
Josephson effect, this expression does not contain explicitely the tunneling
matrix element. Contrary to the ordinary Josephson effect, this equation is
non-linear. Solving it numerically along with Eq. (%
\ref{dS/dt_AC}) in the case where the magnetic field vanishes on one side
of the junction, one finds that the angle between $\mathbf{S}_{L}$ and $%
\mathbf{S}_{R}$ behaves as a sine-like function of time. The presence of an
additionnal constant magnetic field $\mathbf{B}$ throughout the system adds
a beat to this sine-like behavior. The gyromagnetic ratio $g$ is material
dependent. No such non-universal constant appears in the ordinary Josephson
effect. The above discussion of the AC effect can be
transposed for FM by replacing the staggered magnetic moment by the
uniform one.


In summary, we found a Josephson-like equilibrium current of the staggered
magnetic moment through a tunnel junction between two itinerant
AF with non-colinear staggered magnetic moments. In analogy with the
ferromagnetic case, this current is proportional to $\mathbf{S}
_{R}\times \mathbf{S}_{L}$ where $\mathbf{S}_{L}$ and $\mathbf{S}_{R}$ are the staggered magnetic moments on either sides of the junction. The effect comes from tunneling of coherent spin-one neutral
particle-hole pairs with a net momentum and no net spin projection in the order parameter direction. There are important differences with the ordinary
Josephson effect coming both from the different coupling of the
electromagnetic fields and from the non-abelian nature of the broken
symmetry. It would be interesting to perform magnetic resonnance experiments
to detect the effects we predicted.

\textit{Acknowledgments} We are grateful to M.B. Paranjape and R.B.
MacKenzie for discussions that stimulated our interest in this problem and
to A.V. Andreev, A. Blais, D. Bergeron, C. Bourbonnais, R. C\^{o}t\'{e}, P.
Fournier and D. S\'{e}n\'{e}chal for insightful comments. This work was
partially supported by FQRNT (Qu\'{e}bec), by the Tier I Canada Research
Chair Program (A.-M.S.T.) and CIFAR (A.-M.S.T.).


\begin{thebibliography}{99}
\bibitem{half}We limit ourselves to nearest neighbour hopping to simplify the expression of the transformation amplitudes, but the main result of this paper, namely formulae (\ref{mainresult1}) and (\ref{mainresult2}) for the equilibrium current of the staggered magnetic moment, are also valid for an extended model. Dissipationless spin current will occur if U is large enough that there is a gap to quasi-particle excitations.
\end{thebibliography}
\end{document}